\def\@chapapp{Chapter}
\def\chapapp{\@chapapp}

\def\mychapter#1#2{
   \def\@chaphead##1{#1}
   \def\@schaphead##1{#2}
}

\def\@makechapterhead#1{ \vspace*{\chaptopsep} { \parindent 0pt \raggedright
 \ifnum \c@secnumdepth >\m@ne \@chaphead{#1} \else \@schaphead{#1} \fi
 \par \nobreak \vskip \chapaftersep } }

\def\@makeschapterhead#1{ \vspace*{\chaptopsep} { \parindent 0pt \raggedright
 \@schaphead{#1}\par
 \nobreak \vskip \chapaftersep } }

\newlength{\chaptopsep}
\newlength{\chapaftersep}
\mychapter{\LARGE \sc \thechapter. #1}{\LARGE \sc #1}

\def\@startsection#1#2#3#4#5#6{\if@noskipsec \leavevmode \fi
   \par \@tempskipa #4\relax
   \@afterindenttrue
   \ifdim \@tempskipa <\z@ \@tempskipa -\@tempskipa \@afterindentfalse\fi
   \if@nobreak \everypar{}\else
     \addpenalty{\@secpenalty}\addvspace{\@tempskipa}\fi \@ifstar
     {\@ssect{#3}{#4}{#5}{#1}}{\@dblarg{\@sect{#1}{#2}{#3}{#4}{#5}{#6}}}}

\def\@sect#1#2#3#4#5#6[#7]#8{\ifnum #2>\c@secnumdepth
     \def\@svsec{}\else
     \refstepcounter{#1}
     \edef\@svsec{\expandafter\relax\csname @pre#1\endcsname}\fi
     \@tempskipa #5\relax
      \ifdim \@tempskipa>\z@
        \begingroup \expandafter\relax\csname @#1style\endcsname\relax
          \@hangfrom{\hskip #3\relax\@svsec}{\interlinepenalty \@M #8\par}
        \endgroup
       \csname #1mark\endcsname{#7}\addcontentsline
         {toc}{#1}{\ifnum #2>\c@secnumdepth \else
                      \protect\numberline{\csname the#1\endcsname}\fi
                    #7}\else
        \def\@svsechd{%
            \expandafter\relax\csname @#1style\endcsname\relax
            \hskip #3\@svsec #8\csname #1mark\endcsname
                      {#7}\addcontentsline
                           {toc}{#1}{\ifnum #2>\c@secnumdepth \else
                             \protect\numberline{\csname the#1\endcsname}\fi
                       #7}}\fi
     \@xsect{#5}}

\def\@ssect#1#2#3#4#5{\@tempskipa #3\relax
   \ifdim \@tempskipa>\z@
     \begingroup \expandafter\relax\csname @#4style\endcsname\relax
     \@hangfrom{\hskip #1}{\interlinepenalty \@M #5\par}\endgroup
   \else \def\@svsechd{#4\hskip #1\relax #5}\fi
    \@xsect{#3}}

\def\mysection#1#2#3{
     \expandafter\def\csname @#1style\endcsname{#2}
     \expandafter\def\csname @pre#1\endcsname{#3}
}

\mysection{section}{\large\bf}{\thesection.~}
\mysection{subsection}{\normalsize\bf}{\thesubsection.~}
\mysection{subsubsection}{\normalsize\bf}{\thesubsubsection.~}
\mysection{paragraph}{\normalsize\bf}{\theparagraph.}
\mysection{subparagraph}{\normalsize\bf}{\thesubparagraph.}

\def\@begintheorem#1#2{\sl \trivlist
   \item[\hskip \labelsep{\bf #1\ #2\thmcounterend}]}
\def\@opargbegintheorem#1#2#3{\sl \trivlist
      \item[\hskip \labelsep{\bf #1\ #2\ (#3)\thmcounterend}]}
\def\thmcounterend{.}

\def\numberbysection{\renewcommand{\thesection}{\arabic{section}}
                     \renewcommand{\theequation}{\thesection.\arabic{equation}}
                     \@addtoreset{equation}{section}
                     \renewcommand{\thetheorem}{\thesection.\arabic{theorem}}
                     \@addtoreset{theorem}{section}
                     \renewcommand{\thefigure}{\thesection.\arabic{figure}}
                     \@addtoreset{figure}{section}
                     \renewcommand{\thetable}{\thesection.\arabic{table}}
                     \@addtoreset{table}{section}}

\def\numberbysubsection{\renewcommand{\thesection}{\arabic{section}}
              \renewcommand{\thesubsection}{\arabic{subsection}}
              \renewcommand{\theequation}{\thesubsection.\arabic{equation}}
              \@addtoreset{equation}{subsection}
              \renewcommand{\thetheorem}{\thesubsection.\arabic{theorem}}
              \@addtoreset{theorem}{subsection}
              \renewcommand{\thefigure}{\thesubsection.\arabic{figure}}
              \@addtoreset{figure}{subsection}
              \renewcommand{\thetable}{\thesubsection.\arabic{table}}
              \@addtoreset{table}{subsection}}

\@definecounter{theorem}

\def\theenumi{\arabic{enumi}}

\def\theenumii{\arabic{enumii}}
\def\p@enumii{\theenumi.}

\def\theenumiii{\arabic{enumiii}}
\def\p@enumiii{\theenumi.\theenumii.}

\def\p@enumiv{\p@enumiii.\theenumiii}

\def\outline{\ifnum \@enumdepth >3 \@toodeep\else
      \advance\@enumdepth \@ne
      \edef\@enumctr{enum\romannumeral\the\@enumdepth}\list
      {\csname label\@enumctr\endcsname}{\usecounter
        {\@enumctr}\def\makelabel##1{\hss\llap{##1}}
         \parsep \z@ \itemsep \z@
         \ifnum \@enumdepth > 1 \topsep \z@ \fi}\fi}

\def\lhead#1{\gdef\@lhead{#1}\gdef\@erhead{#1}}
\def\lfoot#1{\gdef\@lfoot{#1}\gdef\@erfoot{#1}}
\def\chead#1{\gdef\@chead{#1}\gdef\@echead{#1}}
\def\cfoot#1{\gdef\@cfoot{#1}\gdef\@ecfoot{#1}}
\def\rhead#1{\gdef\@rhead{#1}\gdef\@elhead{#1}}
\def\rfoot#1{\gdef\@rfoot{#1}\gdef\@elfoot{#1}}

\def\@lhead{} \def\@lfoot{}
\def\@chead{} \def\@cfoot{}
\def\@rhead{} \def\@rfoot{}

\def\elhead#1{\gdef\@elhead{#1}} \def\elfoot#1{\gdef\@elfoot{#1}}
\def\echead#1{\gdef\@echead{#1}} \def\ecfoot#1{\gdef\@ecfoot{#1}}
\def\erhead#1{\gdef\@erhead{#1}} \def\erfoot#1{\gdef\@erfoot{#1}}

\def\@elhead{} \def\@elfoot{}
\def\@echead{} \def\@ecfoot{}
\def\@erhead{} \def\@erfoot{}

\def\@threepart#1#2#3{\rlap{#1} \hfil {#2} \hfil \llap{#3}}

\def\ps@threepartheadings
    {
    \def\@oddhead{\@threepart{\@lhead}{\@chead}{\@rhead}}
    \def\@oddfoot{\@threepart{\@lfoot}{\@cfoot}{\@rfoot}}

    \if@twoside
	\def\@evenhead{\@threepart{\@elhead}{\@echead}{\@erhead}}
	\def\@evenfoot{\@threepart{\@elfoot}{\@ecfoot}{\@erfoot}}
    \else
	\def\@evenhead{\@threepart{\@lhead}{\@chead}{\@rhead}}
	\def\@evenfoot{\@threepart{\@lfoot}{\@cfoot}{\@rfoot}}
    \fi
    }

\ps@threepartheadings

\def\underhead{
 \addtolength{\headheight}{\underheadsep}
 \def\@outputpage{\begingroup\catcode`\ =10 \if@specialpage
     \global\@specialpagefalse\@nameuse{ps@\@specialstyle}\fi
     \if@twoside
       \ifodd\count\z@ \let\@thehead\@oddhead \let\@thefoot\@oddfoot
            \let\@themargin\oddsidemargin
          \else \let\@thehead\@evenhead
          \let\@thefoot\@evenfoot \let\@themargin\evensidemargin
     \fi\fi
     \shipout
     \vbox{\normalsize \baselineskip\z@ \lineskip\z@
           \vskip \topmargin \moveright\@themargin
           \vbox{\setbox\@tempboxa
                   \vbox to\headheight{\vfil \hbox to\textwidth{\@thehead}
                         \vskip \underheadsep
                          \if@nounderhead\global\@nounderheadfalse
                                 \hbox to\textwidth{}
                          \else  \hbox to\textwidth{\hrulefill}\fi}
                 \dp\@tempboxa\z@
                 \box\@tempboxa
                 \vskip \headsep
                 \box\@outputbox
                 \baselineskip\footskip
                 \hbox to\textwidth{\@thefoot}}}\global\@colht\textheight
           \endgroup\stepcounter{page}\let\firstmark\botmark}

}
\def\ps@empty{\@nounderheadtrue
              \let\@mkboth\@gobbletwo\def\@oddhead{}\def\@oddfoot{}%
\def\@evenhead{}\def\@evenfoot{}}

\def\ps@plain{\@nounderheadtrue
     \let\@mkboth\@gobbletwo
     \def\@oddhead{}\def\@oddfoot{\rm\hfil\thepage
     \hfil}\def\@evenhead{}\let\@evenfoot\@oddfoot}

\newif\if@nounderhead \@nounderheadfalse
\def\nounderhead{\@nounderheadtrue}

\newlength{\underheadsep}
\def\tighttoc{\def\l@section{\@dottedtocline{1}{0em}{1.4em}}}

\def\symbolnote#1#2{\let\@thefootn=\thefootnote%
\renewcommand{\thefootnote}{\fnsymbol{footnote}}%
\footnotemark[#1]%
\footnotetext[#1]{#2}%
\let\thefootnote=\@thefootn
}

\def\symbolmark#1{\let\@thefootn=\thefootnote%
\renewcommand{\thefootnote}{\fnsymbol{footnote}}%
\footnotemark[#1]%
\let\thefootnote=\@thefootn
}

\def\symboltext#1#2{\let\@thefootn=\thefootnote%
\renewcommand{\thefootnote}{\fnsymbol{footnote}}%
\footnotetext[#1]{#2}%
\let\thefootnote=\@thefootn
}

\def\big#1{{\hbox{$\left#1\vcenter to1.428\ht\strutbox{}\right.\n@space$}}}
\def\Big#1{{\hbox{$\left#1\vcenter to2.142\ht\strutbox{}\right.\n@space$}}}
\def\bigg#1{{\hbox{$\left#1\vcenter to2.857\ht\strutbox{}\right.\n@space$}}}
\def\Bigg#1{{\hbox{$\left#1\vcenter to3.571\ht\strutbox{}\right.\n@space$}}}

\def\docspecials{\do\ \do\$\do\&%
  \do\#\do\^\do\^^K\do\_\do\^^A\do\%\do\~}
{\catcode`\/=0\catcode`\\=12/xdef/@bsl{\}}
\def\literatim#1{\trivlist \item[]\if@minipage\else\vskip\parskip\fi
\leftskip\@totalleftmargin\rightskip\z@
\parindent\z@\parfillskip\@flushglue\parskip\z@
\def\@nll{}\def\@arg{#1}
\ifx\@nll\@arg\def\@newcc{}\else\def\@newcc{\catcode`#1=14}\fi
\let\@=\@bsl
\let\@dollar=$
\let\@amper=&
\let\circumflex=\^
\@tempswafalse \def\par{\if@tempswa\hbox{}\fi\@tempswatrue\@@par}
\obeylines \tt \catcode``=13 \@noligs \let\do\@makeother \docspecials
\let\$=\@dollar
\let\&=\@amper
\let\_=\sb
\let\^=\sp
\frenchspacing\@vobeyspaces\@newcc}

\documentstyle[11pt,jeep]{report}

\typeout{(TCILATEX Macros for Scientific Word 1.1 <18 May 93>.)}
\typeout{(AMSTeX text macro commented out for compatibility with LaTeX.}
\typeout{AMSTEXT style option to be used instead. Alterations marked nad}
\typeout{nad@phys.canterbury.ac.nz 18 Nov 93.}

\makeatletter
\let\DOTSI\relax
\def\RIfM@{\relax\ifmmode}%
\def\FN@{\futurelet\next}%
\newcount\intno@
\def\iint{\DOTSI\intno@\tw@\FN@\ints@}%
\def\iiint{\DOTSI\intno@\thr@@\FN@\ints@}%
\def\iiiint{\DOTSI\intno@4 \FN@\ints@}%
\def\idotsint{\DOTSI\intno@\z@\FN@\ints@}%
\def\ints@{\findlimits@\ints@@}%
\newif\iflimtoken@
\newif\iflimits@
\def\findlimits@{\limtoken@true\ifx\next\limits\limits@true
 \else\ifx\next\nolimits\limits@false\else
 \limtoken@false\ifx\ilimits@\nolimits\limits@false\else
 \ifinner\limits@false\else\limits@true\fi\fi\fi\fi}%
\def\multint@{\int\ifnum\intno@=\z@\intdots@                                
 \else\intkern@\fi                                                          
 \ifnum\intno@>\tw@\int\intkern@\fi                                         
 \ifnum\intno@>\thr@@\int\intkern@\fi                                       
 \int}
\def\multintlimits@{\intop\ifnum\intno@=\z@\intdots@\else\intkern@\fi
 \ifnum\intno@>\tw@\intop\intkern@\fi
 \ifnum\intno@>\thr@@\intop\intkern@\fi\intop}%
\def\intic@{\mathchoice{\hskip.5em}{\hskip.4em}{\hskip.4em}{\hskip.4em}}%
\def\negintic@{\mathchoice
 {\hskip-.5em}{\hskip-.4em}{\hskip-.4em}{\hskip-.4em}}%
\def\ints@@{\iflimtoken@                                                    
 \def\ints@@@{\iflimits@\negintic@\mathop{\intic@\multintlimits@}\limits    
  \else\multint@\nolimits\fi                                                
  \eat@}
 \else                                                                      
 \def\ints@@@{\iflimits@\negintic@
  \mathop{\intic@\multintlimits@}\limits\else
  \multint@\nolimits\fi}\fi\ints@@@}%
\def\intkern@{\mathchoice{\!\!\!}{\!\!}{\!\!}{\!\!}}%
\def\plaincdots@{\mathinner{\cdotp\cdotp\cdotp}}%
\def\intdots@{\mathchoice{\plaincdots@}%
 {{\cdotp}\mkern1.5mu{\cdotp}\mkern1.5mu{\cdotp}}%
 {{\cdotp}\mkern1mu{\cdotp}\mkern1mu{\cdotp}}%
 {{\cdotp}\mkern1mu{\cdotp}\mkern1mu{\cdotp}}}%
\def\Let@{\relax\iffalse{\fi\let\\=\cr\iffalse}\fi}%
\def\vspace@{\def\vspace##1{\crcr\noalign{\vskip##1\relax}}}%
\def\multilimits@{\bgroup\vspace@\Let@
 \baselineskip\fontdimen10 \scriptfont\tw@
 \advance\baselineskip\fontdimen12 \scriptfont\tw@
 \lineskip\thr@@\fontdimen8 \scriptfont\thr@@
 \lineskiplimit\lineskip
 \vbox\bgroup\ialign\bgroup\hfil$\m@th\scriptstyle{##}$\hfil\crcr}%
\def\Sb{_\multilimits@}%
\def\endSb{\crcr\egroup\egroup\egroup}%
\def\Sp{^\multilimits@}%

\newdimen\ex@
\ex@.2326ex
\def\rightarrowfill@#1{$#1\m@th\mathord-\mkern-6mu\cleaders
 \hbox{$#1\mkern-2mu\mathord-\mkern-2mu$}\hfill
 \mkern-6mu\mathord\rightarrow$}%
\def\leftarrowfill@#1{$#1\m@th\mathord\leftarrow\mkern-6mu\cleaders
 \hbox{$#1\mkern-2mu\mathord-\mkern-2mu$}\hfill\mkern-6mu\mathord-$}%
\def\leftrightarrowfill@#1{$#1\m@th\mathord\leftarrow\mkern-6mu\cleaders
 \hbox{$#1\mkern-2mu\mathord-\mkern-2mu$}\hfill
 \mkern-6mu\mathord\rightarrow$}%
\def\overrightarrow{\mathpalette\overrightarrow@}%
\def\overrightarrow@#1#2{\vbox{\ialign{##\crcr\rightarrowfill@#1\crcr
 \noalign{\kern-\ex@\nointerlineskip}$\m@th\hfil#1#2\hfil$\crcr}}}%

\def\overleftarrow{\mathpalette\overleftarrow@}%
\def\overleftarrow@#1#2{\vbox{\ialign{##\crcr\leftarrowfill@#1\crcr
 \noalign{\kern-\ex@\nointerlineskip}$\m@th\hfil#1#2\hfil$\crcr}}}%
\def\overleftrightarrow{\mathpalette\overleftrightarrow@}%
\def\overleftrightarrow@#1#2{\vbox{\ialign{##\crcr\leftrightarrowfill@#1\crcr
 \noalign{\kern-\ex@\nointerlineskip}$\m@th\hfil#1#2\hfil$\crcr}}}%
\def\underrightarrow{\mathpalette\underrightarrow@}%
\def\underrightarrow@#1#2{\vtop{\ialign{##\crcr$\m@th\hfil#1#2\hfil$\crcr
 \noalign{\nointerlineskip}\rightarrowfill@#1\crcr}}}%

\def\underleftarrow{\mathpalette\underleftarrow@}%
\def\underleftarrow@#1#2{\vtop{\ialign{##\crcr$\m@th\hfil#1#2\hfil$\crcr
 \noalign{\nointerlineskip}\leftarrowfill@#1\crcr}}}%
\def\underleftrightarrow{\mathpalette\underleftrightarrow@}%
\def\underleftrightarrow@#1#2{\vtop{\ialign{##\crcr$\m@th\hfil#1#2\hfil$\crcr
 \noalign{\nointerlineskip}\leftrightarrowfill@#1\crcr}}}%
\def\tfrac#1#2{{\textstyle {#1 \over #2}}}%
\newcount\GRAPHICSTYPE
\GRAPHICSTYPE=\z@
\def\GRAPHICSPS#1{%
 \ifnum\GRAPHICSTYPE=\@ne language "PS", include "#1"\else ps: #1\fi
 }%
\def\graffile#1#2#3#4{%
 \leavevmode\raise -#4 \hbox{%
  \raise #3 \hbox{\rule{0.003in}{0.003in}\special{#1}}%
  }%
 {\raise -#4 \hbox to #2 {\vrule height#3 width\z@ depth\z@\hfil}}%
 }%
\def\draftbox#1#2#3#4{%
 \leavevmode\raise -#4 \hbox{%
  \frame{\rlap{\protect\tiny #1}\hbox to #2%
   {\vrule height#3 width\z@ depth\z@\hfil}%
  }%
 }%
}%
\newcount\draft
\draft=\z@
\def\GRAPHIC#1#2#3#4#5{%
 \ifnum\draft=\@ne \draftbox{#2}{#3}{#4}{#5}%
  \else \graffile{#1}{#3}{#4}{#5}%
  \fi
 }%
\def\addtoLaTeXparams#1{\edef\LaTeXparams{\LaTeXparams #1}}%
\def\doFRAMEparams#1{\readFRAMEparams#1\end}%
\def\readFRAMEparams#1{%
 \ifx#1\end%
  \let\next=\relax
  \else
  \ifx#1i\dispkind=\z@\fi
  \ifx#1d\dispkind=\@ne\fi
  \ifx#1f\dispkind=\tw@\fi
  \ifx#1t\addtoLaTeXparams{t}\fi
  \ifx#1b\addtoLaTeXparams{b}\fi
  \ifx#1p\addtoLaTeXparams{p}\fi
  \ifx#1h\addtoLaTeXparams{h}\fi
  \let\next=\readFRAMEparams
  \fi
 \next
 }%
\def\IFRAME#1#2#3#4#5{\GRAPHIC{#5}{#4}{#1}{#2}{#3}}%
\def\DFRAME#1#2#3#4{%
 \begin{center}\GRAPHIC{#4}{#3}{#1}{#2}{\z@}\end{center}%
 }%
\def\FFRAME#1#2#3#4#5#6#7{%
 \begin{figure}[#1]%
  \begin{center}\GRAPHIC{#7}{#6}{#2}{#3}{\z@}\end{center}%
  \caption{\label{#5}#4}%
  \end{figure}%
 }%
\newcount\dispkind%
\def\FRAME#1#2#3#4#5#6#7#8{%
 \def\LaTeXparams{}%
 \dispkind=\z@
 \def\LaTeXparams{}%
 \doFRAMEparams{#1}%
 \ifnum\dispkind=\z@\IFRAME{#2}{#3}{#4}{#7}{#8}\else
  \ifnum\dispkind=\@ne\DFRAME{#2}{#3}{#7}{#8}\else
   \ifnum\dispkind=\tw@
    \edef\@tempa{\noexpand\FFRAME{\LaTeXparams}}%
    \@tempa{#2}{#3}{#5}{#6}{#7}{#8}%
    \fi
   \fi
  \fi
 }%
\long\def\QQQ#1#2{\long\expandafter\def\csname#1\endcsname{#2}}%
\def\QTP#1{}%
\long\def\QQA#1#2{}%
\def\QTR#1#2{{\csname#1\endcsname #2}}
\long\def\TeXButton#1#2{#2}%
\def\EXPAND#1[#2]#3{}%
\def\NOEXPAND#1[#2]#3{}%
\def\LaTeXparent#1{}%
\def\QTagDef#1#2#3{}%
\def\QQfnmark#1{\footnotemark}

\@ifundefined{INDEX}{\def\INDEX#1#2{}{}}{}%
\@ifundefined{SUBINDEX}{\def\SUBINDEX#1#2#3{}{}{}}{}%
\def\initial#1{\bigbreak{\raggedright\large\bf #1}\kern 2\p@\penalty3000}%
\@ifundefined{abstract}{%
 \def\abstract{%
  \if@twocolumn
   \section*{Abstract (Not appropriate in this style!)}%
   \else \small
   \begin{center}{\bf Abstract\vspace{-.5em}\vspace{\z@}}\end{center}%
   \quotation
   \fi
  }%
 }{%
 }%
\@ifundefined{endabstract}{\def\endabstract
  {\if@twocolumn\else\endquotation\fi}}{}%
\@ifundefined{maketitle}{\def\maketitle#1{}}{}%
\@ifundefined{affiliation}{\def\affiliation#1{}}{}%
\@ifundefined{proof}{}{}%
\@ifundefined{endproof}{}{}%
\@ifundefined{newfield}{\def\newfield#1#2{}}{}%
\@ifundefined{chapter}{\def\chapter#1{\par(Chapter head:)#1\par }%
 \newcount\c@chapter}{}%
\@ifundefined{part}{\def\part#1{\par(Part head:)#1\par }}{}%
\@ifundefined{section}{\def\section#1{\par(Section head:)#1\par }}{}%
\@ifundefined{subsection}{\def\subsection#1%
 {\par(Subsection head:)#1\par }}{}%
\@ifundefined{subsubsection}{\def\subsubsection#1%
 {\par(Subsubsection head:)#1\par }}{}%
\@ifundefined{paragraph}{\def\paragraph#1%
 {\par(Subsubsubsection head:)#1\par }}{}%
\@ifundefined{subparagraph}{\def\subparagraph#1%
 {\par(Subsubsubsubsection head:)#1\par }}{}%
\@ifundefined{therefore}{}{}%
\@ifundefined{backepsilon}{}{}%
\@ifundefined{yen}{}{}%
\@ifundefined{registered}{\def\registered{\relax\ifmmode{}\r@gistered
                                                \else$\m@th\r@gistered$\fi}%
 \def\r@gistered{^{\ooalign
  {\hfil\raise.07ex\hbox{$\scriptstyle\rm\text{R}$}\hfil\crcr
  \mathhexbox20D}}}}{}%
\@ifundefined{Eth}{}{}%
\@ifundefined{eth}{}{}%
\@ifundefined{Thorn}{}{}%
\@ifundefined{thorn}{}{}%
\@ifundefined{degree}{}{}%
\def\BibTeX{{\rm B\kern-.05em{\sc i\kern-.025em b}\kern-.08em
    T\kern-.1667em\lower.7ex\hbox{E}\kern-.125emX}}%
\newdimen\theight
\def\Column{%
 \vadjust{\setbox\z@=\hbox{\scriptsize\quad\quad tcol}%
  \theight=\ht\z@\advance\theight by \dp\z@\advance\theight by \lineskip
  \kern -\theight \vbox to \theight{%
   \rightline{\rlap{\box\z@}}%
   \vss
   }%
  }%
 }%
\def\qed{%
 \ifhmode\unskip\nobreak\fi\ifmmode\ifinner\else\hskip5\p@\fi\fi
 \hbox{\hskip5\p@\vrule width4\p@ height6\p@ depth1.5\p@\hskip\p@}%
 }%
\def\miss{\hbox{\vrule height2\p@ width 2\p@ depth\z@}}%
\def\tcol#1{{\baselineskip=6\p@ \vcenter{#1}} \Column}  %
\makeatother
\renewcommand{\theequation}{\arabic{equation}}
\parskip=0pt plus 1pt
\parindent=16pt
\hfuzz=2pt
\vfuzz=4pt
\hsize=31pc
\vsize=55 truepc
\setlength{\topmargin}{-1.5cm}
\setlength{\textheight}{23.0cm}
\addtolength{\hoffset}{-0.5cm}
\addtolength{\textwidth}{1cm}
\lhead{\small Parabosons \& Jahn-Teller systems}
\chead{\thepage}
\rhead{\small \today}
\thispagestyle{plain}

\QQQ{Language}{
British English
}

\begin{document}

\TeXButton{title}
{
{\Large Parabosons versus supersymmetry in Jahn-Teller Systems}

\bigskip

\bigskip

}

\TeXButton{author,address,abstract}
{
Edward D. Savage \& G E Stedman,

\bigskip

Department of Physics and Astronomy, University of Canterbury, Private Bag
4800,\\
Christchurch, New Zealand

\bigskip

\bigskip

The applications of parabosons by Schmutz (1980) and of supersymmetry
 by Jarvis {\em et al.} (1984) in Jahn-Teller systems are compared and
contrasted.
Although a  parasupersymmetric
Jahn-Teller system has not yet been identified, the method of  Schmutz
is used here to show that the $E  \times \epsilon $
 supersymmetric Jahn-Teller Hamiltonian can be written in terms of
paraboson operators.

\bigskip

\bigskip

}

\TeXButton{today}{\today}

\newpage\

\TeXButton{1. Intro}{{\large 1. Introduction}}

Jahn-Teller systems are interesting candidates for nonrelativistic
applications of supersymmetry in quantum mechanics because of the degeneracy
of the fermion states, the existing fermion-boson interactions and the
well-established tradition for applying higher group symmetries to reveal
approximate underlying symmetries (for example Pooler and O'Brien 1977, Judd
1982, Stedman 1983).

The work of Schmutz (1980) on parabosons and that of Jarvis {\em et al.}
(1984) on supersymmetry, respectively, in $E\otimes \epsilon $ Jahn-Teller
Hamiltonians have superficially common features. We show that although the
differences in these approaches are fundamental, and do not allow the
identification of a parasupersymmetric Jahn-Teller system at this stage, the
anharmonic terms introduced by Jarvis {\em et al. }to achieve supersymmetry
may be given elegant representation using paraboson operators.

We take the Hamiltonian to be\ $H\equiv H_l+H_e+H_c+H_a$ where $H_e,H_l$ are
the unperturbed Hamiltonians of the $N-$ fold degenerate electronic system
and of a harmonic oscillator with the same degeneracy (so that supersymmetry
is possible), $H_c$ is the fermion-boson coupling term and $H_a$ represents
anharmonic phonon coupling. In the schemes of Jarvis {\em et al.} (1984) it
is vital that anharmonic (boson-boson) couplings in $H_c$ be present to act
as the supersymmetric counterparts of the fermion-boson couplings under
fermion-boson transmutations. Since physical systems will certainly possess
some anharmonicity, this supersymmetric model is expected to be at least as
realistic as those assuming harmonic couplings when discussing higher
symmetry in Jahn-Teller systems.

In the $E\otimes \epsilon $ system a doubly degenerate vibrational mode (of
symmetry $\epsilon $ in say the group O) is vibronically coupled to a
twofold degenerate (E) electronic level ($H_e=${\bf 1}$)$. The fermion-boson
coupling has the form $H_c\equiv \TeXButton{sigma}
{\mbox{{\boldmath$\sigma$}}}_z\phi _1+\TeXButton{sigma}
{\mbox{{\boldmath$\sigma$}}}_x\phi _2$ where $\phi _i=b_i+b_i^{\dagger },$ $%
b_i,$ $f_i$ are annihilation operators for boson mode and fermion state $i,$
respectively, and \thinspace $\TeXButton{sigma}{\mbox{{\boldmath$\sigma$}}}%
_z $, $\TeXButton{sigma}{\mbox{{\boldmath$\sigma$}}}_x$ are the usual Pauli
matrices. We shall write {\bf b, f} for the associated column matrices $%
\left( b_i\right) ,\left( f_i\right) .$\medskip\

\TeXButton{2. SUSY JT}{{\large 2. Supersymmetry in Jahn-Teller systems}}%
\label{sec2}

We now review and adapt the results of the Jarvis {\em et al.} formalism.
The generator of supersymmetric transformations is the supercharge $S\equiv $%
{\bf f}$^{\dagger }{\cdot }\TeXButton{beta}{\mbox{{\boldmath$\beta$}}}$,
where $\TeXButton{beta}{\mbox{{\boldmath$\beta$}}}\equiv \exp G(%
\TeXButton{phi}{\mbox{{\boldmath$\phi$}}}){\bf b}\exp (-G(\TeXButton{phi}
{\mbox{{\boldmath$\phi$}}}))$, $\TeXButton{phi}{\mbox{{\boldmath$\phi$}}}%
\equiv \{\phi _i\},i=1,2,$ and $G$ is any real differentiable function of $%
\TeXButton{phi}{\mbox{{\boldmath$\phi$}}}$. Therefore: $\left[
\overleftarrow{\TeXButton{beta}{\mbox{{\boldmath$\beta$}}}},\overrightarrow{%
\TeXButton{beta}{\mbox{{\boldmath$\beta$}}}}\right] =0,\qquad \left[
\overleftarrow{\TeXButton{beta}{\mbox{{\boldmath$\beta$}}}},\overrightarrow{%
\text{{\bf f}}^{\dagger }}\right] =\left[ \overleftarrow{\TeXButton{beta}
{\mbox{{\boldmath$\beta$}}}},\overrightarrow{\text{{\bf f}}}\right] =0,$
where the left and right arrows indicate row and column labels,
respectively. $S$ is nilpotent, and the Hamiltonian $H=\{S,S^{\dagger }\}$
is necessarily supersymmetric (Witten 1982, Blockley{\em \ et al.} 1985),
Stedman 1985).

We may expand $\TeXButton{beta}{\mbox{{\boldmath$\beta$}}}$ in terms of
repeated commutators and use the result: $\left[ G(\TeXButton{phi}
{\mbox{{\boldmath$\phi$}}}),b_i^{(\pm )}\right] =\pm G^{(i)}(\TeXButton{phi}
{\mbox{{\boldmath$\phi$}}})$, where the superscript $i$ denotes a partial
derivative with respect to $\phi _i$. Since any two functions of $%
\TeXButton{phi}{\mbox{{\boldmath$\phi$}}}$ commute we have $\TeXButton{beta}
{\mbox{{\boldmath$\beta$}}}={\bf b}-{\bf G}^{\prime }(\TeXButton{phi}
{\mbox{{\boldmath$\phi$}}})$. It follows that $\left[ \overleftarrow{%
\TeXButton{beta}{\mbox{{\boldmath$\beta$}}}},\overrightarrow{\TeXButton{beta}
{\mbox{{\boldmath$\beta$}}}^{\dagger }}\right] =${\bf 1}$-2${\bf G}$^{\prime
\prime }(\TeXButton{phi}{\mbox{{\boldmath$\phi$}}})$. The Hamiltonian
becomes
\begin{equation}
\begin{array}[t]{lll}
H & = & \text{{\bf f}}^{\dagger }\left[ \overleftarrow{\TeXButton{beta}
{\mbox{{\boldmath$\beta$}}}},\ \overrightarrow{\TeXButton{beta}
{\mbox{{\boldmath$\beta$}}}^{\dagger }}\right] \text{{\bf f}}+%
\TeXButton{beta}{\mbox{{\boldmath$\beta$}}}^{\dagger }\TeXButton{beta}
{\mbox{{\boldmath$\beta$}}}, \\  & = & \text{{\bf f}}^{\dagger }\text{{\bf f}%
}+\text{{\bf b}}^{\dagger }\text{{\bf b}}-2\text{{\bf f}}^{\dagger }{\bf G}%
^{\prime \prime }(\TeXButton{phi}{\mbox{{\boldmath$\phi$}}})\text{{\bf f}}-%
\text{{\bf G}}^{\prime }(\TeXButton{phi}{\mbox{{\boldmath$\phi$}}})\text{%
{\bf b}}-\text{{\bf b}}^{\dagger }\text{{\bf G}}^{\prime }(\TeXButton{phi}
{\mbox{{\boldmath$\phi$}}})+\text{{\bf G}}^{\prime }(\TeXButton{phi}
{\mbox{{\boldmath$\phi$}}})^2.
\end{array}
\end{equation}
This may be written as $H=H_e+H_l+H_c+H_a,$ where $H_c=$ $-2${\bf f}$%
^{\dagger }${\bf G}$^{\prime \prime }(\TeXButton{phi}
{\mbox{{\boldmath$\phi$}}})${\bf f} and
\begin{equation}
\begin{array}[t]{lll}
H_a & = & -
{\bf G}^{\prime }(\TeXButton{phi}{\mbox{{\boldmath$\phi$}}}){\bf b}-{\bf b}%
^{\dagger }{\bf G}^{\prime }(\TeXButton{phi}{\mbox{{\boldmath$\phi$}}})+{\bf %
G}^{\prime }(\TeXButton{phi}{\mbox{{\boldmath$\phi$}}})^2, \\  & = & tr{\bf G%
}^{\prime \prime }(\TeXButton{phi}{\mbox{{\boldmath$\phi$}}})-({\bf G}%
^{\prime }(\TeXButton{phi}{\mbox{{\boldmath$\phi$}}}))^T\TeXButton{phi}
{\mbox{{\boldmath$\phi$}}}+{\bf G}^{\prime }(\TeXButton{phi}
{\mbox{{\boldmath$\phi$}}})^2.\
\end{array}
\end{equation}
Jarvis {\it et al.} point out that each such term in $H$ is guaranteed to be
invariant under the point group if the fermions and bosons transform in the
same manner and if $G(\TeXButton{phi}{\mbox{{\boldmath$\phi$}}})$ is an
invariant function; this follows since {\ $\TeXButton{beta}
{\mbox{{\boldmath$\beta$}}}\sim {\bf b}\sim {\bf f}$ }and a contraction such
as ${\bf f}^{\dagger }\TeXButton{beta}{\mbox{{\boldmath$\beta$}}}$ is then
invariant under the point group.

For the $E\otimes \epsilon $ Jahn-Teller system, Jarvis {\it et al. }(1984)
choose a D$_4\supset $ D$_2$ subgroup basis so that $\TeXButton{phi}
{\mbox{{\boldmath$\phi$}}}=\left( \phi _1,\phi _2\right) ^T$ $\sim
((3z^2-r^2)/\surd 3,x^2-y^2)^T$. The only quadratic, cubic and quartic
invariants that can be constructed from $\TeXButton{phi}
{\mbox{{\boldmath$\phi$}}}${\bf \ }are $I_2\equiv (\phi _1^2+\phi _2^2)$, $%
I_3\equiv (\phi _1^3-3\phi _1\phi _2^2)$, and $I_4\equiv I_2^2$. Also ${\bf %
\Phi }\equiv (\phi _1^2-\phi _2{}^2,-2\phi _1\phi _2)^T$ transforms as $%
\TeXButton{phi}{\mbox{{\boldmath$\phi$}}}$. If $G_E\equiv -{\frac 13}\alpha
I_3$, then $\TeXButton{beta}{\mbox{{\boldmath$\beta$}}}={\bf b}+\alpha {\bf %
\Phi }$ and
\begin{equation}
\begin{array}{rcl}
H_c & = & -2{\bf f}^{\dagger }\left(
\begin{array}{cc}
-2\alpha \phi _1 & 2\alpha \phi _2 \\
2\alpha \phi _2 & 2\alpha \phi _1
\end{array}
\right)
{\bf f} \\  & = & 4\alpha (\phi _1(f_1^{\dagger }f_1-f_2^{\dagger }f_2)-\phi
_2(f_1^{\dagger }f_2-f_2^{\dagger }f_1)) \\
& = & 4\alpha {\bf f}^{\dagger }(\phi _1{\bf \sigma }_z-\phi _2{\bf \sigma }%
_x){\bf f}.
\end{array}
\end{equation}
$G_E$ generates a mixture of cubic and quartic anharmonicity: $H_a=\alpha
I_3+\alpha ^2I_4$. Thus,
\begin{equation}
H_{JS}=H_2+H_e+H_c+H_a={\bf b}^{\dagger }{\bf b}+{\bf f}^{\dagger }{\bf f}%
+4\alpha {\bf f}^{\dagger }(\phi _1{\bf \sigma }_z-\phi _2{\bf \sigma }_x)%
{\bf f}+\alpha I_3+\alpha ^2I_4.
\end{equation}
On projecting out the fermion operators $H_{JS}\rightarrow {\bf H}_{JS}$ by
the relation $H_{JS}\equiv {\bf f}^{\dagger }{\bf H}_{JS}{\bf f}$, we find
\begin{equation}
\begin{array}{lll}
{\bf H}_{JS} & = & \left[
\overleftarrow{\TeXButton{beta}{\mbox{{\boldmath$\beta$}}}},\overrightarrow{%
\TeXButton{beta}{\mbox{{\boldmath$\beta$}}}}^{\dagger }\right] +\left(
\TeXButton{beta}{\mbox{{\boldmath$\beta$}}}^{\dagger }\TeXButton{beta}
{\mbox{{\boldmath$\beta$}}}\right) {\bf 1}, \\  & = & ({\bf b}^{\dagger }%
{\bf b}+1+\alpha I_3+\alpha ^2I_4){\bf 1}+4\alpha (\phi _1{\bf \sigma }%
_z-\phi _2{\bf \sigma }_x).
\end{array}
\end{equation}

\medskip\

\TeXButton{3. PBose JT}{{\large 3. Parabosons in Jahn-Teller systems}}

Similarly, we briefly review and adapt the representation by Schmutz (1980)
of an $E\otimes \epsilon $ Jahn-Teller system in terms of displaced parabose
oscillators (we use $\alpha =\lambda /4,$ $\phi _2\rightarrow -\phi _2$ ).
Schmutz begins with the $E\otimes \epsilon $ Jahn-Teller Hamiltonian ${\bf H}%
_2+{\bf H}_e+{\bf H}_c$:
\begin{equation}
{\bf H}_S=({\bf b}^{\dagger }{\bf b}+1){\bf 1}+4\alpha \left( \phi _1{\bf %
\sigma }_z-\phi _2{\bf \sigma }_x\right) =\left( {\bf b}^{\dagger }{\bf b}%
\right) {\bf 1}+\left[ \overleftarrow{\TeXButton{beta}
{\mbox{{\boldmath$\beta$}}}},\overrightarrow{\TeXButton{beta}
{\mbox{{\boldmath$\beta$}}}^{\dagger }}\right] .
\end{equation}
Note the omission of anharmonicity. The operator $\Gamma _i\equiv \exp (i\pi
b_i^{\dagger }b_i)$ and has the useful properties $\Gamma _i^{\dagger
}=\Gamma _i^{-1}$ ,  $\{\Gamma _i,b_i\}=0,$ $\Gamma _i|n_i\rangle
=(-1)^{n_i}|n_i\rangle $; in addition, since $\Gamma _i^2$ has expectation
value unity in any space of definite (integer) number, and commutes with all
operators in the theory, we can take $\Gamma _i=\Gamma _i^{\dagger }=\Gamma
_i^{-1}$ in all relations. A derived unitary operator ${\bf U}_1$
diagonalizes ${\bf H}_S$:
\begin{equation}
{\bf U}_i=\frac 1{\sqrt{2}}\left(
\begin{array}{cc}
1 & \Gamma _i \\
1 & -\Gamma _i
\end{array}
\right) ,\;{\bf H}_S{}^{\prime }={\bf U}_1{\bf H}_S{\bf U}_1^{\dagger
}=\left(
\begin{array}{cc}
H_{-} & 0 \\
0 & H_{+}
\end{array}
\right) ;
\end{equation}
\begin{equation}
H_\eta \equiv {\bf b}^{_{\dagger }}{\bf b}+1+4\alpha (\phi _1+\eta \phi
_2\Gamma _1),.
\end{equation}
where $\eta =\pm $. The operators $a_i,a_i^{\dagger }$ where $a_i\equiv
b_i\Gamma _{3-i}$, $i=1,2,$ obey boson commutation relations amongst
themselves (as do $b_i,b_i^{\dagger }$), but each has zero anticommutator
with each of $b_i,b_i^{\dagger }.$ ${\bf A}\equiv \left( A_{+},A_{-}\right)
^T={\bf U}_1{\bf b,}$ $A_\eta \equiv (b_1+\eta a_2)/\surd 2.$ These
operators satisfy the trilinear commutation relations characteristic of all $%
p=2$ parabosonic operators (where $\zeta $, like $\eta $, is either + or -)
(see for related material Green 1953, Greenberg and Messiah 1965, Rubakov
and Spironodov 1988, Beckers and Debergh 1990a,b, Bardakci 1992):
\begin{equation}
[\{A_\eta ,A_\eta ^{\dagger }\},A_{-\eta }]=-A_{-\eta },\;[\{A_\eta ,A_\eta
\},A_\eta ^{\dagger }]=2A_\eta ,\;[\{A_\eta ,A_\eta \},A_\zeta ]=0.
\end{equation}
Other such relations follow by hermitian conjugation and also by the
generalized Jacobi identity. $A_{\pm }$ may be defined by the relations $%
[N_\zeta ,A_\eta ]=-A_\eta ,$ where $N_\zeta \equiv \{A_\zeta ,A_\zeta
^{\dagger }\}.$ In the two-dimensional case, $N_{+}=N_{-}\equiv N$ and is
the number operator for the system. The unperturbed Hamiltonian of the two
dimensional harmonic oscillator, $H_2={\bf b}^{\dagger }{\bf b}+1,$ may be
written as $H_2=N={\bf A}^{\dagger }{\bf A}+1.$ $H_\eta $ can therefore be
expressed as $H_\eta =N_\eta ^{^{\prime }}-16\alpha ^2,$ where $N_\zeta
^{^{\prime }}\equiv \{A_\zeta ^{^{\prime }},A_\zeta ^{^{\prime }\dagger }\}$
and $A_\eta ^{^{\prime }}\equiv A_\eta +2\surd 2\alpha $ so that $N_\zeta
^{^{\prime }}=N+2\sqrt{2}\alpha \left( A_\zeta +A_\zeta ^{\dagger }\right)
+8\alpha ^2.$

Hence the Hamiltonians $H_\eta $ are identical with those of displaced
parabose oscillators ($A_\eta $ being the parabose operator and $2\surd
2\alpha $ its displacement). Under the unitary transformation ${\bf U}_1$ in
which ${\bf H}_S\rightarrow {\bf H}_S^{^{\prime }},$ ${\bf f\rightarrow f}%
^{\prime }$ (which preserves the Fermi anticommutation relations)${\bf ,}$
then $H_S={\bf f}^{\prime {\bf \dagger }}{\bf H}_S^{^{\prime }}{\bf f}%
^{\prime }{\bf ,}$and parabosonic expressions are obtained in the
Hamiltonian. The Schmutz diagonalisation process may thus be viewed as a
result of this unitary symmetry of the formalism.

\medskip\

\TeXButton{4. Amalgm}
{{\large 4. Action of Schmutz transformations in the Jarvis {\it et al.}
Hamiltonian}}

An obvious question is: what is the action of the Schmutz unitary
diagonalising matrix ${\bf U}_1$ on the Jarvis {\em et al. }Hamiltonian $%
{\bf H}_{JS}$? Let ${\bf H}_{JS}\rightarrow {\bf H}_{JS}^{^{\prime }},$ $%
{\bf f\rightarrow f}^{\prime }$ and $\TeXButton{beta}
{\mbox{{\boldmath$\beta$}}}\rightarrow \TeXButton{beta}
{\mbox{{\boldmath$\beta$}}}^{\prime }$ under ${\bf U}_1${\bf .} We obtain $%
{\bf H}_{JS}^{\prime }={\bf H}_S^{\prime }+\alpha I_3\TeXButton{sigma}
{\mbox{{\boldmath$\sigma$}}}{_x}+\alpha ^2I_4{\bf 1}.$ Since $H_{JS}={\bf f}%
^{\prime \dagger }{\bf H}_{JS}^{^{\prime }}{\bf f}^{\prime },$ this may be
regarded as the original Hamiltonian in a unitarily transformed fermion
basis. We note also that $S={\bf f}^{\prime \dagger }\TeXButton{beta}
{\mbox{{\boldmath$\beta$}}}^{\prime },$ where $\TeXButton{beta}
{\mbox{{\boldmath$\beta$}}}^{\prime }\equiv {\bf A}+\alpha {\bf P}_1$ and $%
{\bf P}_1\equiv \surd 2\left(
\begin{array}{c}
C_{-}C_{+} \\
C_{+}C_{-}
\end{array}
\right) $ with $C_\zeta \equiv A_\zeta +A_\zeta ^{\dagger }.$ Hence $S$ also
involves paraboson operators, and the (supersymmetric) Hamiltonian%
$$
H_{JS}=\left( {\bf f}^{\prime \dagger }\TeXButton{beta}
{\mbox{{\boldmath$\beta$}}}^{\prime }\right) \left( \TeXButton{beta}
{\mbox{{\boldmath$\beta$}}}^{\prime ^{\dagger }}{\bf f}^{\prime }\right)
+\left( \TeXButton{beta}{\mbox{{\boldmath$\beta$}}}^{\prime ^{\dagger }}{\bf %
f}^{\prime }\right) \left( {\bf f}^{\prime \dagger }\TeXButton{beta}
{\mbox{{\boldmath$\beta$}}}^{\prime }\right) ={\bf f}^{\prime \dagger }{\bf H%
}_{JS}^{^{\prime }}{\bf f}^{\prime }.
$$
Since ${\bf b}^{\dagger }{\bf b}={\bf A}^{\dagger }{\bf A},$
\begin{equation}
\TeXButton{beta}{\mbox{{\boldmath$\beta$}}}^{^{\prime }\dagger }%
\TeXButton{beta}{\mbox{{\boldmath$\beta$}}}^{\prime }={\bf A}^{\dagger }{\bf %
A}+\alpha ({\bf A}^{\dagger }{\bf P}_1+{\bf P}_1^{\dagger }{\bf A})+\alpha ^2%
{\bf P}_1^{\dagger }{\bf P}_1,
\end{equation}
$I_3=\sqrt{2}({\bf A}^{\dagger }{\bf P}_1+{\bf P}_1^{\dagger }{\bf A}%
)=C_{+}C_{-}C_{+},$ $I_4={\bf P}_1^{\dagger }{\bf P}_1$ $%
=\{C_{+}C_{-},C_{-}C_{+}\},$ and so the Jahn-Teller Hamiltonian may be
written
\begin{equation}
H_{JS}={\bf f}^{\prime \dagger }\left( {\bf H}_S+\alpha ({\bf P}_1^{\dagger }%
{\bf A}+{\bf A}{^{\dagger }}{\bf P}_1)\TeXButton{sigma}
{\mbox{{\boldmath$\sigma$}}}_x+\alpha ^2{\bf P}_1^{\dagger }{\bf P}_1\right)
{\bf f}^{\prime }.
\end{equation}
At this point we simply follow Schmutz's method for rendering such a
Hamiltonian diagonal, using the fermion transformation associated with ${\bf %
U}_2$; and, as in the work of Schmutz, the effect is to highlight further
the paraboson operators. If ${\bf U\equiv U}_2{\bf U}_1,$ ${\bf H}%
_{JS}\rightarrow {\bf H}_{JS}^{^{\prime \prime }},{\bf f\rightarrow f}%
^{\prime \prime }$ and $\TeXButton{beta}{\mbox{{\boldmath$\beta$}}}%
\rightarrow \TeXButton{beta}{\mbox{{\boldmath$\beta$}}}^{\prime \prime }$
under ${\bf U}$ then:
\begin{equation}
{\bf H}_{JS}^{^{\prime \prime }}=({\bf b}^{\dagger }{\bf b}+1+4\alpha (\phi
_1-\phi _2\Gamma _1)){\bf 1}+\alpha I_3\Gamma _2\TeXButton{sigma}
{\mbox{{\boldmath$\sigma$}}}_z+\alpha ^2I_4{\bf 1}=\left(
\begin{array}{cc}
H_{JS+} & 0 \\
0 & H_{JS-}
\end{array}
\right) ,
\end{equation}
where
\begin{equation}
\begin{array}[t]{rcl}
H_{JS\eta } & = & {\bf b}^{\dagger }{\bf b}+1+4\alpha (\phi _1-\phi _2\Gamma
_1)+\eta \alpha I_3\Gamma _2+\alpha ^2I_4 \\  & = & (N_{-}^{^{\prime
}}-16\alpha ^2+\alpha ^2I_4){\bf 1}+\eta \alpha I_3\Gamma _2
\end{array}
\end{equation}
or, in terms of paraboson operators:
$$
{\bf H}_{JS}=(N_{-}^{^{\prime }}-16\alpha ^2){\bf 1}+\alpha ({\bf P}%
_1^{\dagger }{\bf A}+{\bf A}^{\dagger }{\bf P}_1)\Gamma _2\TeXButton{sigma}
{\mbox{{\boldmath$\sigma$}}}_z+\alpha ^2\left( {\bf P}_1^{\dagger }{\bf P}%
_1\right) {\bf 1}.
$$
Hence
\begin{equation}
H_{JS}={\bf f}^{\prime \prime ^{\dagger }}({\bf U}\left[ \overleftarrow{%
\TeXButton{beta}{\mbox{{\boldmath$\beta$}}}},\overrightarrow{\TeXButton{beta}
{\mbox{{\boldmath$\beta$}}}^{\dagger }}\right] {\bf U}^{\dagger }+{\bf U}%
\TeXButton{beta}{\mbox{{\boldmath$\beta$}}}^{\prime \prime \ \dagger }%
\TeXButton{beta}{\mbox{{\boldmath$\beta$}}}^{\prime \prime }{\bf U}^{\dagger
}){\bf f}^{\prime \prime }.
\end{equation}

It is instructive to write in the last term
\begin{equation}
\TeXButton{beta}{\mbox{{\boldmath$\beta$}}}^{\prime \prime \ \dagger }%
\TeXButton{beta}{\mbox{{\boldmath$\beta$}}}^{\prime \prime }=\TeXButton{beta}
{\mbox{{\boldmath$\beta$}}}^{\dagger }(\Gamma _2{\bf U}_2)^{\dagger }(\Gamma
_2{\bf U}_2)\TeXButton{beta}{\mbox{{\boldmath$\beta$}}}={\bf b}^{\dagger }%
{\bf b}+\alpha ({\bf B}^{\dagger }{\bf P}_2+{\bf P}_2^{\dagger }{\bf B}%
)\Gamma _2+\alpha ^2{\bf P}_2^{\dagger }{\bf P}_2,
\end{equation}
where ${\bf B}\equiv \left(
\begin{array}{c}
B_{+} \\
B_{-}
\end{array}
\right) =\Gamma _2{\bf U}_2{\bf b},\;{\bf P}_2\equiv {\bf U}_2{\bf \Phi }%
=\surd \bar 2\left(
\begin{array}{c}
D_{+}D_{-} \\
D_{-}D_{+}
\end{array}
\right) ,$ $D_\zeta \equiv B_\zeta +B_\zeta ^{\dagger },$ $B_\zeta \equiv
\left( b_1\Gamma _2+\zeta b_2\right) /\sqrt{2}.$ $\left\{ B_\zeta \right\} $
are therefore para-boson operators (i.e. obey a mixture of commutation and
anticommutation relations). In addition ${\bf B}^{\dagger }{\bf B=b}%
^{\dagger }{\bf b}={\bf A}^{\dagger }{\bf A},$ $I_3=({\bf B}^{\dagger }{\bf P%
}_2+{\bf P}_2^{\dagger }{\bf B})\Gamma _2,$ $I_3\Gamma _2=$ $\sqrt{2}\left(
D_{+}D_{-}D_{+}\text{ }+D_{-}D_{+}D_{-}\right) ,$ $I_4={\bf P}_2^{\dagger }%
{\bf P}_2$ $=2\left\{ D_{+}D_{-},D_{-}D_{+}\right\} ,$ so that

\begin{equation}
{\bf U}\TeXButton{beta}{\mbox{{\boldmath$\beta$}}}^{\ \dagger }%
\TeXButton{beta}{\mbox{{\boldmath$\beta$}}}{\bf U}^{\dagger }={\bf A}%
^{\dagger }{\bf A}+\alpha ({\bf B}^{\dagger }{\bf P}_2+{\bf P}_2^{\dagger }%
{\bf B})\Gamma _2+\alpha ^2{\bf P}_2^{\dagger }{\bf P}_2,
\end{equation}
and thus

\begin{equation}
{\bf H}_{JS}({\bf A,B})=(N_{-}^{^{\prime }}-16\alpha ^2){\bf 1}+\alpha ({\bf %
P}_2^{\dagger }{\bf B}+{\bf B}^{\dagger }{\bf P}_2)\TeXButton{sigma}
{\mbox{{\boldmath$\sigma$}}}_z+\alpha ^2\left( {\bf P}_2^{\dagger }{\bf P}%
_2\right) {\bf 1}.
\end{equation}
This is a diagonal Hamiltonian expressed entirely in terms of parabosons.

Remarkably the anharmonic terms are very amenable to expression in terms of
paraboson operators. The cubic and quartic anharmonicity invariants $I_3$
and $I_4$ have in fact a far more elegant relationship when so expressed;
the symmetry between the plus and minus parabosons is manifest.

\TeXButton{5 PSUSY}{{\large 5. Parasupersymmetry?}}

All of the former analysis is confined to standard fermi-bose supersymmetry,
whose generator and spectrum are those reported in Jarvis {\em et al.}
(1984). However the above analysis is suggestive of a role for
parasupersymmetry in Jahn-Teller theory.

The usual approach, however, is to use parafermi-bose supersymmetry (Jarvis
1978, Rubakov {\em et al.} 1988, Beckers and Debergh 1990a). This can lead
to spectra with threefold degeneracies. Such a possibility of alternative
higher symmetries would continue and enhance the above-mentioned tradition
for applying higher group symmetries in Jahn-Teller systems.

Following the first example of Rubakov {\em et al.,} we might search for
parasupersymmetry using the paracharge%
$$
Q=\left(
\begin{array}{ccc}
0 & 0 & 0 \\
p+iW_1 & 0 & 0 \\
0 & p+iW_2 & 0
\end{array}
\right) ,
$$
leading to a Hamiltonian of the form
$$
H=\tfrac 12p^2+W_1^2+W_2^2+\tfrac 13\left( W_1^{\prime }-W_2^{\prime
}\right) +\tfrac 13\left(
\begin{array}{ccc}
2W_1^{\prime }+W_2^{\prime } & 0 & 0 \\
0 & W_2^{\prime }-W_1^{\prime } & 0 \\
0 & 0 & -W_1^{\prime }-2W_2^{\prime }
\end{array}
\right)
$$
For this to replicate, say the T$\,\times \epsilon $ Jahn-Teller system, we
need to identify this interaction by an appropriate choice of the
superpotentials $W_1^{},W_2.$  The T$\,\times \epsilon $ Jahn-Teller system
has an interaction, when diagonalised, of the form:%
$$
\left(
\begin{array}{ccc}
\sqrt{\phi _1^2+\phi _2^2} & 0 & 0 \\
0 & -\sqrt{\phi _1^2+\phi _2^2} & 0 \\
0 & 0 & 0
\end{array}
\right) .
$$
Hence we would have to identify%
$$
2W_1^{\prime }+W_2^{\prime }=W_1^{\prime }-W_2^{\prime }=3W_1^{\prime }/2=%
\sqrt{\phi _1^2+\phi _2^2},\,\,W_1^{\prime }=-2W_2^{\prime }.
$$
While these relations are algebraically consistent, each potential must
contain both $\phi _1$and $\phi _2$; in addition, the further conditions
required by Rubakov {\em et al}. give the unlikely requirement that $%
W_2^{\prime }\left( W_2^{}+2W_1\right) =3W_2^{\prime \prime }.$ This
approach therefore seems unpromising. Nevertheless the paraboson link
established here may help to indicate a better direction for studying
possible realisations of parasupersymmetric systems.

\TeXButton{Ackn}{{\large Acknowledgements}}

We are grateful to U Merkel for bringing the work of Schmutz to our
attention.

\newpage\

\TeXButton{Refs}{{\large References}}

Bardakci K, 1992 {\em Nucl. Phys.} {\em B} {\bf 369} 461-477

Beckers J and Debergh N, 1990a {\em Nucl. Phys. B }{\bf 340} 767-776

--------- 1990b\ {\em J. Math. Phys.} {\bf 31 1513-23}

Blockley C\ A and Stedman G\ E{\it ,} 1985 {\em Eur. J. Phys.} {\bf 6}
218-224

Green H\ S, 1953 {\em Phys. Rev.} {\bf 90} 270-3

Greenberg O\ W and Messiah A\ M\ L, 1965{\it \ }{\em Phys. Rev.B} {\bf 138}
1555-67

Jarvis P\ D\ 1978 {\em Aust. J. Phys. }{\bf 31}, 461-469

Jarvis P\ D and Stedman G\ E, 1984 {\em J. Phys. A: Math. Gen.} {\bf 17}
757-776

Judd B\ R, 1982 {\em Physica} {\bf 114A }19-27

Pooler, D\ R and O'Brien M C M, 1977 {\em J. Phys. C: Solid State Phys.}{\bf %
\ 10} 3769-91

Rubakov V\ A and Spiridonov V\ P, 1988 {\em Mod. Phys. Lett. A} {\bf 3}
1337-1347

Schmutz M, 1980 {\em Physica} {\bf 101}A 1-21

Stedman G\ E,\ 1983 {\em Eur. J. Phys.} {\bf 4} 156-61

--------- 1985{\it \ }{\em Eur. J. Phys}. {\bf 6} 225-231

Witten E, 1982{\it \ }{\em Nucl. Phys.} B {\bf 202} 253--256

\end{document}